\documentclass[         %
aps,                    
prd,                    
showpacs,               
nofootinbib,            
twocolumn,             %
showkeys,               %
preprintnumbers,        %
floatfix]               
{revtex4}               

\usepackage{graphicx,longtable}
\usepackage{epsfig}

\begin{document}

\title{Does the neutrino magnetic moment have an impact on solar 
neutrino physics?}

\author{        A.~B. Balantekin}
\email{         baha@nucth.physics.wisc.edu}
\affiliation{Institut de Physique Nucl\'eaire, F-91406 Orsay cedex, 
France \\and Department of Physics, University of Wisconsin, 
Madison, Wisconsin 53706 USA\footnote{Permanent Address}}
\author{        C. Volpe}
\email{volpe@ipno.in2p3.fr}
\affiliation{Institut de Physique Nucl\'eaire, F-91406 Orsay cedex,
France}

\date{\today}
\begin{abstract}
Solar neutrino observations coupled with the recent KamLAND data 
suggest that spin-flavor precession scenario does not play a major role 
in neutrino propagation in the solar matter. We provide approximate 
analytical formulas and numerical results 
to estimate the contribution of the spin-flavor 
precession, if any, to the electron neutrino survival probability when 
the magnetic moment and magnetic field combination is small. 
\end{abstract}
\medskip
\preprint{}
\maketitle

\section{Introduction}

In the earlier days of the solar neutrino research activities 
one of the more speculative solutions proposed to resolve the puzzle 
of missing neutrinos invoked the interaction of the neutrino magnetic 
moment with the solar magnetic fields. 
Although initial attempts \cite{Cisneros:1970nq} 
ignored matter effects, eventually the combined effect of matter and 
magnetic fields was brought out \cite{Lim:1987tk,Akhmedov:uk}. 
Simultaneous presence of a large neutrino magnetic moment, magnetic 
field combination and neutrino flavor mixing can give rise to two 
additional resonances besides the MSW 
resonance \cite{Wolfenstein:1977ue,Mikheev:wj}. Initial numerical 
calculations \cite{Balantekin:1990jg} using the resonant spin-flavor 
precession scheme were carried out before the gallium experiments 
were completed. These calculations hinted a solution of the solar 
neutrino problem with the neutrino parameters in the LOW region 
provided that there is transition magnetic moment as large as 
$10^{-11} \mu_B$ and a magnetic field of the order of $10^5$ G. 
With this solution count rate at the gallium detectors would be 
significantly reduced. Even though gallium experiments ruled out 
this particular solution variants of the spin-flavor precession 
solution to the solar neutrino problem continue to be 
investigated by many researchers. (A representative set of the 
recent work is given in Refs. 
\cite{Guzzo:1998sb,Akhmedov:2002mf,Friedland:2002pg,Kang:2004tx}).

In the meantime experimental data has increasingly disfavored the 
spin-flavor precession solution to the solar neutrino problem. 
Earlier reports of an anticorrelation between the solar magnetic 
activity and solar neutrino capture rate at the Homestake detector 
\cite{Davis:cj} was a prime motivation for considering the 
magnetic field effects. An analysis of the Super-Kamiokande data 
rules out such an anticorrelation \cite{Yoo:2003rc}. (For a counter 
argument, however see Ref. \cite{Sturrock:2004hx}). Also, if the 
neutrinos are of Majorana type, such a scenario would produce 
solar antineutrinos \cite{Raghavan:1991em,Akhmedov:uk1}. 
Both the Super-Kamiokande \cite{Gando:2002ub} and  
KamLAND collaborations \cite{Eguchi:2003gg} failed to find 
any evidence for solar antineutrinos. Finally a global fit of 
all solar neutrino experiments {\em without} using the 
spin-flavor precession solution (see e.g. Ref. 
\cite{Balantekin:2003jm}) was confirmed by the KamLAND 
experiment. 

On the other hand our knowledge of both the solar magnetic fields 
and neutrino magnetic moments has been improving. (At the very least 
we now know that that neutrinos, since they are definitely massive, 
have magnetic moments the magnitude of which depends on the physics 
beyond the Standard Model). Hence it may be worthwhile to revisit 
the spin-flavor precession mechanism and knowing it is not the 
dominant mechanism, explore what implications it may still have. 

\section{Status on Solar Magnetic Fields and $\mu_{\nu}$}

Neither the magnetic pressure in the core nor the impact of the 
structure of the magnetic field on the stellar dynamics are usually 
taken into account in the Standard Solar Model 
\cite{Bahcall:2000nu,Brun:1998qk}. Within the uncertainties of the 
nuclear physics input \cite{Adelberger:1998qm} Standard Solar Model 
agrees well not only with the neutrino observations 
 but also with the 
helioseismological observations of the sound speed profile
\cite{Bahcall:2004fg,Turck,Couvidat:ba}. 
Direct measurements of the magnetic fields in the radiative zone 
with acoustic modes are not possible, however 
even with a one-dimensional solar model a sizable magnetic field 
would contribute additional pressure. An analysis found that 
a magnetic field greater than $\sim 10^7$ G localized 
at about 0.2 R$_{\odot}$ 
would cause the sound 
speed profile to deviate from the observed values 
\cite{Couvidat:ba}. However this particular calculation does not place 
restrictions on the magnetic field exactly at the center of the Sun or 
even at 0.1 R$_{\odot}$. Similarly 
magnetic field strengths greater than $ \sim 7 \times 10^6$ G 
are not allowed since they could cancel the observed oblateness of 
the Sun, making it spherical and even prolate 
\cite{Friedland:2002is}. 

Helioseismology provides more detailed information 
about the magnetic fields in the convective zone. 
The rotation profile of the Sun is presently known down to about 
0.2 R$_{\odot}$: the rotation of the solar radiative zone is like 
that of a solid body rotating at a constant rate \cite{Couvidat:2003ep}.
Such a rotation profile suggests the existence of a magnetic field  
in the radiative zone.   
The observations 
of the splittings of the solar oscillation frequencies can be used 
to infer the magnetic field. The odd terms in the azimuthal order 
are determined only by the rotation rate in the solar interior. The 
even terms may receive additional contributions from the magnetic 
fields. An analysis of the helioseismic data indicates that rotation 
alone is not sufficient to explain the observed even splitting 
coefficients. Other helioseismic observations are consistent with a
magnetic  
field of $\sim 20$ G at a depth of 30000 km below the solar surface 
\cite{Antia:2000pu}. Similar arguments limit the toroidal magnetic 
field to $< 300$ kG at the bottom of the convective zone. 
On the contrary, at present, there is no direct helioseismic evidence 
for the presence or absence of sizable magnetic fields in the
radiative zone.  

As far as the neutrino magnetic moment is concerned, best present
direct upper limits come from reactor experiments, i.e. 
$\mu_{\nu} < 1.0-1.3 \times 10^{-10} \mu_B$ at $90 \%$ C.L.  
\cite{Daraktchieva:2003dr,texono} which
improve previous bounds \cite{savannah,Vogel:iv,kurchatov,rovno},
as well as from the recent Super-Kamiokande solar data
which has given the limit of
$ < 1.5 \times 10^{-10} \mu_B$ at $90 \%$ C.L. 
\cite{Beacom:1999wx}. 
Combining recent solar neutrino experiments with the KamLAND data 
yields a limit of $ < 1.1 \times 10^{-10} \mu_B$ at $90 \%$ C.L..  
From astrophysical considerations, an indirect upper limit in the range
of $10^{-11}- 10^{-12} \mu_B$ have been obtained \cite{Raffelt:wa},
the exact limits being model dependent. 
New experiments are now under study which would lower the direct
limits down to the level of a few $\times 10^{-12}$  
using a static tritium source 
\cite{ioannis,mamont,McLaughlin:2003yg} while the use of
low energy beta-beams might lower it by about one order of magnitude
\cite{McLaughlin:2003yg,lownu}.

\section{Spin-Flavor Precession Formalism and Results}

The evolution of the chiral components of two flavors of 
neutrinos is described by \cite{Lim:1987tk}
\begin{equation}
\label{1}
i \frac{d}{dt} 
\left(\matrix{\nu^{(L)}_e \cr \nu^{(L)}_{\mu} \cr  \nu^{(R)}_e \cr  
\nu^{(R)}_{\mu} \cr } \right) =
\left( \matrix{ H^{(L)} & BM^{\dagger} \cr
                B M  &    H^{(R)} } \right)
\left(\matrix{\nu^{(L)}_e \cr \nu^{(L)}_{\mu} \cr  \nu^{(R)}_e \cr  
\nu^{(R)}_{\mu} \cr } \right). 
\end{equation}   
For the Dirac neutrinos one has 
\begin{equation}
\label{2}
H^{(L)} = 
\left( \matrix{ \frac{\delta m^2}{2E} \sin^2 \theta + V_e & 
                      \frac{\delta m^2}{4E} \sin 2\theta  \cr
 \frac{\delta m^2}{4E} \sin 2\theta & 
 \frac{\delta m^2}{2E} \cos^2 \theta + V_{\mu} \cr } \right),
\end{equation}
and $H^{(R)}$ is given by setting $V_e$ and $V_{\mu}$ equal to 
zero in Eq. (\ref{2}). For the Majorana neutrinos in Eq. (\ref{1}) 
one write down for the left-handed component
\begin{equation}
\label{3}
H^{(L)} = 
\left( \matrix{ V_e & 
                      \frac{\delta m^2}{4E} \sin 2\theta  \cr
 \frac{\delta m^2}{4E} \sin 2\theta & 
 \frac{\delta m^2}{2E} \cos 2\theta + V_{\mu} \cr } \right).
\end{equation}
For the Majorana neutrinos the right-handed part of the Hamiltonian,
$H^{(R)}$, is given by replacing $V_e$ and $V_{\mu}$ in Eq. 
(\ref{3}) by $-V_e$ and $-V_{\mu}$, respectively. In these equations 
the matter potentials are
\begin{equation}
\label{4a}
V_e = \frac{G_F}{\sqrt{2}} (2 N_e - N_n),
\end{equation}
and
\begin{equation}
\label{4b}
V_{\mu} = - \frac{G_F}{\sqrt{2}} N_n,
\end{equation}
where $G_F$ is the Fermi constant of the weak interactions, $N_e$ is 
the electron density, and $N_n$ is the neutron density. In the above 
equations for the Dirac neutrinos a general magnetic moment matrix is 
possible:
\begin{equation}
\label{5}
M = \left( \matrix{ \mu_{ee} & \mu_{e \mu} \cr
           \mu_{\mu e} & \mu_{\mu \mu} \cr} \right) . 
\end{equation}
For the Majorana neutrinos the diagonal components of Eq. (\ref{5}) 
vanish and the off-diagonal components are related by $- \mu_{e \mu} = 
\mu_{\mu e} \equiv \mu $.

In this scenario there are several resonances. In addition to the 
standard MSW resonance ($\nu^{(L)}_e \rightarrow \nu^{(L)}_{\mu}$) 
that takes place where the condition 
\begin{equation}
\label{6a}
\sqrt{2} G_F N_e = \frac{\delta m^2}{2E_{\nu}} \cos 2\theta
\end{equation} 
is satisfied in the Sun. For the left-handed electron neutrinos that 
are produced the core of the Sun a second, spin-flavor precession, 
resonance ($\nu^{(L)}_e \rightarrow \nu^{(R)}_{\mu}$) 
is possible. For the Dirac neutrinos it takes place where 
the condition 
\begin{equation}
\label{6b}
\frac{G_F}{\sqrt{2}} ( 2 N_e - N_n) = \frac{\delta m^2}{2E_{\nu}} 
\cos 2\theta
\end{equation} 
is satisfied whereas for Majorana neutrinos it is where the condition 
\begin{equation}
\label{6c}
\sqrt{2} G_F ( N_e - N_n) = \frac{\delta m^2}{2E_{\nu}} \cos 2\theta
\end{equation} 
is satisfied. This resonance converts a left-handed electron neutrino 
into a right-handed (sterile) muon neutrino for the Dirac case and 
into a muon anti-neutrino in the Majorana case. (In principle there is 
another resonance possible for Dirac neutrinos, converting the chirality 
of the electron neutrino, but keeping its flavor the same 
through a diagonal 
moment. But for the neutrinos that also go through the MSW resonance 
-as higher energy solar neutrinos do- this requires a very high neutron 
density, $N_n = 2 N_e$, which is not realized in the Sun). Clearly the 
resonances of (\ref{6b}) and (\ref{6c}) that flip both the chirality 
and the flavor of the electron 
neutrino produced in the nuclear reactions at the solar core take 
place at a higher electron density than the MSW resonance density.
Even though their locations are different for the Dirac and Majorana 
cases, neutrinos need to go through them first. To estimate its location 
we use the approximate expression for the solar electron density 
\cite{bahcall}
\begin{equation}
\label{7}
N_e (r) = 245 \exp ( - 10.54 r / R_{\odot} ) N_A \> {\rm cm}^{-3},
\end{equation}
where $N_A$ is the Avogadro's number. For the solar neutron density we 
use a spline fit to the values given in Ref. \cite{Couvidat:ba}. 
Using the values of $\delta m^2 = 8.2 \times 10^{-5}$ eV$^2$ and 
$\tan^2 \theta = 0.4$, obtained from a 
global analysis of the solar neutrino and most recent KamLAND data 
\cite{:2004mb} we calculate the location of both the spin-flavor and 
MSW resonances. Results for the Majorana neutrinos are given in 
Table I. One can see that since the magnetic field should be present 
at the location of the spin-flavor precession resonance only fields 
at and very near the core play a role.  

\begin{table}[t]
\vspace{-2mm}
\caption{\label{table1} The location of the MSW and spin-flavor 
precession (SFP) 
resonances for Majorana neutrinos. Neutrino energies are given in MeV.  
The $r/R_{\odot}$ value for the location of the resonances are shown.}
\begin{tabular*}{0.48\textwidth}{@{\extracolsep{\fill}}ccc}
\hline\hline
$E_{\nu}$ & SFP  &  MSW \\
\hline
2.50  &  0   & 0.07 \\
3.35 &  0.05  & 0.10 \\
5.00 &  0.10  & 0.13 \\
8.00 &  0.15  & 0.18 \\
13.00 & 0.20  & 0.22 \\
\hline\hline
\end{tabular*}
\end{table}

In this paper we present several approximate formulae for the 
electron-neutrino survival probability in several limiting cases. We 
first consider the case of a small mixing angle (non-adiabatic limit). 
In this case the SFP and MSW resonances are well separated. 
The derivation of the reduction of the electron neutrino survival
probability is presented in detail in the appendix. 
Following Eq. (\ref{aa17})  
the electron neutrino oscillation probability in presence of
a magnetic field can be rewritten as   
\begin{eqnarray}
\label{8}
&& P (\nu_{el} \rightarrow \nu_{eL} , \mu B \neq 0) = 
P (\nu_{el} \rightarrow \nu_{eL} , \mu B = 0 ) \nonumber \\
&\times& 
\exp \left[ - \pi (\mu B) \delta r \right], 
\end{eqnarray}
where the width of the SFP resonance (see Eq.(\ref{aa13}) of the 
appendix) is given as 
\begin{equation}
\label{9-1}
\delta r = \left| \left( \frac{N_e' -N_n'}{N_e -N_n} 
\right)_{\rm at \>\> res.} \right|^{-1} 
\frac{4 \mu B}{\delta m^2 \cos 2 \theta}. 
\end{equation}
Clearly even for the rather large values of the magnetic field 
$B \sim 10^6$ G and the  
magnetic moment of $10^{-11} \> \mu_B$ for a 10 MeV 
neutrino the width of the spin-flavor resonance would be very small, 
i.e. $\left( \frac{\delta r}{R_{\odot}} \right) \sim 0.002$.
It is worth to 
reiterate that the value of the magnetic field at the close vicinity 
of the solar core is not restricted by helioseismology as our 
approximations are no longer valid for very large values of the magnetic 
field.

It is not possible to find an expression similar to Eq. (\ref{8}) for larger 
mixing angles or the adiabatic limit in which case 
the SFP and MSW resonances are 
no longer well-separated. 
However it may be beneficial to 
investigate the unrealistic, but pedagogically instructive limit where the 
neutron density vanishes. In this limit the SFP and MSW resonances 
overlap for any values of the neutrino parameters and magnetic field.  
From Eq. (\ref{a11}) one can calculate  
\begin{eqnarray}
\label{new1}
\frac{d^2}{dt^2} \nu^{(L)}_e &+& \left( \phi^2 + i \frac{d \phi}{dt} +
\Delta^2 + (\mu B)^2 \right) \nu^{(L)}_e \nonumber \\ &+& 
\mu B \sqrt{2} G_F N_n
\nu^{(R)}_{\mu} = 0
\end{eqnarray}
In writing the above equation we assumed that the magnetic field is a
constant. In the limit $N_n$ goes to zero the above
equation becomes an equation for $\nu^{(L)}_e$ only and can easily be solved 
using the semiclassical methods of Ref. \cite{Balantekin:1996ag}. 
For large initial electron densities we obtain 
\begin{equation}
  \label{new2}
   P(\nu_e \rightarrow \nu_e) = \frac{1}{2} - \frac{1}{2}
\cos{2\theta_v}  \left( 1
- 2 P_{\rm hop} \right),
\end{equation}
where the hopping probability is given by
\begin{equation}
\label{new3}
P_{hop} = \exp (- \pi \Omega ),
\end{equation}
with 
\begin{eqnarray}
\label{new4}
\Omega &=& \frac{i}{\pi} 
\int^{r_0^*}_{r_0} dr \frac{\delta m^2}{2 E} \left[ 
\left( \zeta^2(r) - 2\zeta(r)\cos{2\theta_v} +1 \right) \right. 
\nonumber \\ &+& \left. 
(\mu B)^2
\right]^{1/2}.
\end{eqnarray}
where ${r_0^*}$ and ${r_0}$ are the turning points (zeros) of the
integrand. In this expression we introduced the scaled density
\begin{equation}
\label{new5}
\zeta(r) = \frac{2\sqrt{2} G_F N_e(r)}{\delta m^2/E} .
\end{equation}
Using the Taylor series expansion of Eq. (\ref{new4}) for small values 
of $\mu B$ we can write
\begin{eqnarray}
\label{new6}
P_{hop} (\mu B \neq 0) = P_{hop} (\mu B = 0) 
\times \nonumber \\
\exp  \left\{
\frac{i}{\pi} 
\int^{r_0^*}_{r_0} dr \frac{\delta m^2}{2 E} \left[ \frac{(\mu B)^2}{
\sqrt{ \zeta^2(r) - 2\zeta(r)\cos{2\theta_v} +1 }} \right] \right\}.
\end{eqnarray}
For an exponential density, $N_e (r) = n_0 e^{-\alpha r}$, the integral 
above can be calculated to give a hopping reduction factor
\begin{equation}
\label{new7}
\exp \left[ - \frac{\pi}{\alpha} \frac{(\mu B)^2 2 E}{\delta m^2} 
\right].
\end{equation}
For a $10^5$ G magnetic field, a magnetic moment of $10^{-12} \mu_B$,
and $E \sim 10 MeV$, this hopping reduction factor is very small,
$ \sim 10^{-3}$.
For the near-adiabatic limit, not only the hopping reduction factor
is very small, but also the hopping probability itself is significantly
reduced.  Consequently one expects the change in the electron neutrino
survival probability due to a near-allowed value of the neutrino
magnetic moment to be very small. We do not expect using a realistic,
non-zero value of the neutron density to change this conclusion. For
$N_n \neq 0$, however, an analytic expression does not exist.
Therefore, for the case of the observed large mixing angle,
we present results of
a numerical calculation, obtained by solving directly Eq.(24).
Figure 1 shows the electron survival probability
as a function of the electron neutrino energy in presence of a magnetic
field of $10^5$ G and a magnetic moment of $10^{-11} \mu_B$.
In particular a gaussian profile has been taken for the magnetic field.
Note that such results are identical (at a level at least of $10^5$)
to those obtained with the MSW effect only. This result is not changed
if a width twice or five times larger than Eq.(12) is taken.
\begin{figure}[t]
\begin{center}
\epsfig{file=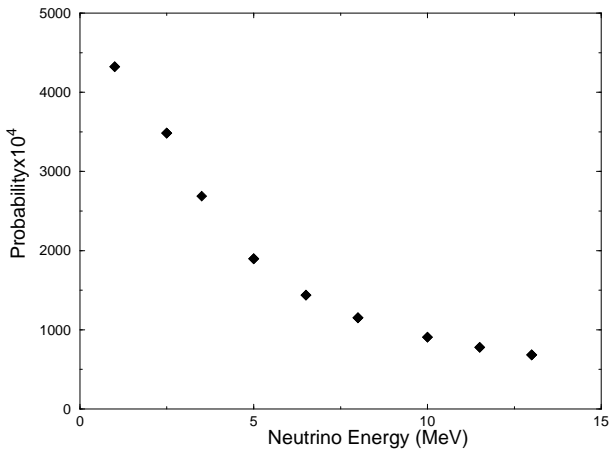,width=6.cm,angle=-90}
\end{center}
\caption{Electron neutrino survival probability 
as a function of neutrino energy, in presence of both the MSW and the
RSF resonances. In particular, a gaussian magnetic field profile is taken,
with a magnetic
field of $10^5$ G and a magnetic moment of $10^{-11} \mu_B$. The location
of the resonances are given in Table I. The results shown are
indistinguishable from
those due to the MSW resonance only, at least at the level of $10^5$.}
\label{fig:prob}
\end{figure}

It was recently argued in Ref. \cite{Miranda:2004nz} that random magnetic 
fields \cite{Loreti:1994ry} can increase the reduction factor by enhancing 
solar antineutrino flux. It is possible to show that this conclusion also 
follows from our approximate expressions. 
We assume a random magnetic field with the 
correlation function \cite{Loreti:1994ry}
\begin{equation}
\label{11}
< B(r) B(r')> = B_0^2 (t) L \delta (r-r')
\end{equation}
where $B_0$ is the average value of the magnetic field and $L$ is the 
correlation length. We rewrite the reduction factor given in 
Eq. (\ref{aa16}) in the form 
\begin{equation}
\label{12}
{\cal R} = 2 \mu^2 \int_0^T dt B(t) e^{iQ(t)} \int_0^t dt' B(t') 
e^{-iQ(t)} ,
\end{equation}
where we defined
\begin{equation}
\label{13}
Q(t) = \int_0^t dt' [ \phi (t') + \kappa (t') ].
\end{equation}
Using Eq. (\ref{11}), Eq. (\ref{12}) can be easily integrated to yield 
\begin{equation}
\label{14}
{\cal R} = 2 \mu^2 L \int_0^T dt B_0^2 (t), 
\end{equation}
which can clearly be very large depending on the chosen average magnetic 
field profile. 

In conclusion, neutrinos emitted by the sun undergo a spin-flavor
precession resonance and then an MSW resonance. In this paper 
we have discussed the conditions to be met to encounter such
resonances and derived analytical formulae for
the reduction of the electron survival probability 
due to solar magnetic fields. Our results show that the coupling of
the neutrino magnetic moment to the solar magnetic field have small effects 
on the
neutrino fluxes.
Such results indicate that future solar neutrino measurements 
could not easily reach the level of precision to pinpoint alternative 
solutions to the solar neutrino deficit than the oscillation 
one confirmed by the recent KamLAND data.

\vspace{.3cm}
We thank S.~Turck-Chieze for useful discussions. 
This  work   was supported in  part  by   the  U.S.  National
Science Foundation Grant No. PHY-0244384 at the
University of  Wisconsin, in  part by  the  University of
Wisconsin Research Committee   with  funds  granted by the
Wisconsin Alumni  Research Foundation, and in part 
by the IN2P3 at IPN-Orsay. ABB thanks the Theoretical 
Physics group at IPN-Orsay for its hospitality during the course 
of this work.

\section*{Appendix: Reduction of the electron neutrino
survival probability in the 
presence of magnetic fields}

In order to obtain the reduction of the electron neutrino survival 
probability due to solar magnetic fields we implement the logarithmic 
perturbation theory (for a description see the Appendix of Ref. 
\cite{Balantekin:1996pp}). Here we show the derivation of the  
probability reduction 
in the case of Majorana neutrinos. A similar derivation can 
be given for the Dirac case. 
Note that the results presented here can also be used to explore
the effect of the magnetic fields on the neutrino 
fluxes produced during the explosion of core-collapse supernovae. 
Since, as described above, there is no resonant 
production of $\nu^{(R)}_e$ in the Sun, we will set its amplitude to 
zero in Eq. (\ref{1}) to obtain
\begin{equation}
\label{a11}
i \frac{d}{dt} 
\left(\matrix{\nu^{(L)}_e \cr \nu^{(L)}_{\mu} \cr   
\nu^{(R)}_{\mu} \cr } \right) =
\left( \matrix{ \phi & \Delta  & \mu B \cr
               \Delta & - \phi & 0 \cr 
               \mu B & 0 & - \kappa } \right)
\left(\matrix{\nu^{(L)}_e \cr \nu^{(L)}_{\mu} \cr    
\nu^{(R)}_{\mu} \cr } \right) 
\end{equation}  
where we defined 
\begin{equation}
\label{a12}
\Delta = \frac{\delta m^2}{4E} \sin 2 \theta, 
\end{equation}
\begin{equation}
\label{a13}
\phi = \frac{1}{\sqrt{2}} G_F N_e -  \frac{\delta m^2}{4E} \cos 
2 \theta,
\end{equation}
and
\begin{equation}
\label{a14}
\kappa = \phi - \sqrt{2} G_F N_n.  
\end{equation}
Introducing
\begin{equation}
\label{a1}
z^{(L)} = \frac{\nu^{(L)}_{\mu}}{\nu^{(L)}_e}, 
\end{equation}
and
\begin{equation}
\label{a2}
z^{(R)} = \frac{\nu^{(R)}_{\mu}}{\nu^{(L)}_e}
\end{equation}
Eq. (\ref{1}) can be rewritten as two coupled nonlinear equations: 
\begin{equation}
\label{a3}
i \frac{dz^{(L)}}{dt} = \Delta - 2 \phi z^{(L)} - \Delta [z^{(L)}]^2 
- \mu B z^{(L)} z^{(R)}, 
\end{equation}
and 
\begin{equation}
\label{a4}
i \frac{dz^{(R)}}{dt} = \mu B - (\kappa + \phi)  z^{(R)}  - 
\Delta z^{(L)} z^{(R)} - \mu B [z^{(R)}]^2.
\end{equation}
From Eqs. (\ref{a3}) and (\ref{a4}) it follows that 
\begin{eqnarray}
\label{a5}
i \frac{d}{dt} &\log& \left( 1 +  |z^{(L)}|^2 + |z^{(R)}|^2 \right)
\nonumber \\
&=& \mu B \left( z^{(R)*} -  z^{(R)} \right) + \Delta \left( z^{(L)*} 
-  z^{(L)} \right). 
\end{eqnarray}
Using the unitarity of the neutrino amplitudes (i.e. $ 1 +  |z^{(L)}|^2 
+ |z^{(R)}|^2 = |\nu^{(L)}_e|^{-2}$) we obtain an {\em exact} 
expression for the electron neutrino amplitude at any location inside or 
outside the Sun:
\begin{eqnarray}
\label{a6}
&&|\nu^{(L)}_e (T)|^2 \nonumber \\ 
&=& \exp \left\{ i \int_0^T dt \left[ \mu B (t) 
\left( z^{(R)*} (t) 
-  z^{(R)} (t) \right) \right]  \right\} \nonumber \\ 
&\times&  \exp \left\{ i \int_0^T dt \left[ 
\Delta \left( z^{(L)*} (t) 
-  z^{(L)} (t) \right) \right] \right\}. 
\end{eqnarray}

Eq. (\ref{a6}), which is an exact result, represents the formulation of  
the N-flavor (or antiflavor) propagation problem in the $SU(N)/SU(N-1) 
\times U(1)$ coset space instead of the usual $SU(N)$ one. To illustrate 
its utility in separating different contributions we 
define the perturbation parameter
\begin{equation}
\label{a7}
g = \frac{\mu B_0}{\Delta} = \frac{4E\mu B_0}{\delta m^2 \sin 2 \theta},
\end{equation}
which we will take to be small. In Eq. (\ref{a7}) $B_0$ is the maximal 
value of the magnetic field. We write the magnetic field in dimensionless 
form
\begin{equation}
\label{a7a}
\beta = B / B_0 . 
\end{equation}
Defining a new variable $\tau = \Delta t$, 
Eqs. (\ref{a3}) and (\ref{a4}) take the form
\begin{equation}
\label{a8}
i \frac{dz^{(L)}}{d\tau} = 1 - 2 \frac{\phi}{\Delta} z^{(L)} - 
[z^{(L)}]^2 - g \beta z^{(L)} z^{(R)}, 
\end{equation}
and 
\begin{equation}
\label{a9}
i \frac{dz^{(R)}}{d\tau} = g \beta - \frac{\kappa + \phi}{\Delta}  
z^{(R)}  - 
z^{(L)} z^{(R)} - g \beta [z^{(R)}]^2.
\end{equation}
These equations need to be solved with the initial condition
$z^{(L)} = 0 = z^{(R)}$ at $t=0$. We consider a perturbative solution of 
the form
\begin{equation}
\label{aa1}
z^{(L)} = z^{(L)}_0 + g z^{(L)}_1 + g^2 z^{(L)}_2 + \cdots ,
\end{equation}
and 
\begin{equation}
\label{aa2}
z^{(R)} = z^{(R)}_0 + g z^{(R)}_1 + g^2 z^{(R)}_2 + \cdots .
\end{equation}
Clearly $z^{(L)}_0$, when substituted into Eq. (\ref{a6}) gives the 
MSW solution and the terms proportional to various powers of $g$  
are the corrections due to the existence of the spin-flavor precession. 

The quantity $z^{(L)}_0$ satisfies the equation
\begin{equation}
\label{aa3}
i \frac{dz^{(L)}_0}{d\tau} = 1 - 2 \frac{\phi}{\Delta} z^{(L)}_0 - 
[z^{(L)}_0]^2 , 
\end{equation}
whereas the evolution of the $z^{(L)}_1$ is given by
\begin{equation}
\label{aa4}
i \frac{dz^{(L)}_1}{d\tau} = - 2 \frac{\phi}{\Delta} z^{(L)}_1 - 
2 z^{(L)}_0 z^{(L)}_1. 
\end{equation}
Eq. (\ref{aa4}) implies that that the quantity 
$z^{(L)}_1$ is a constant times an 
exponential. The only way to satisfy the initial condition is to set 
this multiplicative constant to zero. Hence the lowest order correction 
to  $z^{(L)}$ is  $z^{(L)}_2$, satisfying the equation
\begin{equation}
\label{aa5}
i \frac{dz^{(L)}_2}{d\tau} = - 2 \frac{\phi}{\Delta} z^{(L)}_2 - 
2 z^{(L)}_0 z^{(L)}_2 - \beta z^{(L)}_0 z^{(R)}_1. 
\end{equation}
Similarly, as it is expected on physical grounds, $z^{(R)}_0$ vanishes. 
The lowest order correction to  $z^{(R)}$ is given by $z^{(R)}_1$, which 
satisfies the equation
\begin{equation}
\label{aa6}
i \frac{dz^{(R)}_1}{d\tau} = \beta - \frac{\phi + \kappa}{\Delta} 
z^{(R)}_1 - z^{(L)}_0 z^{(R)}_1. 
\end{equation}
The solution of Eq. (\ref{aa6}) is given by
\begin{eqnarray}
\label{aa7}
&&g z^{(R)}_1 (T) = -i e^{[ i \int_0^T dt' [\phi (t') + \kappa (t') + 
{\Delta} z^{(L)}_0 (t')] ]} \nonumber \\
&\times&  \int_0^T dt \mu B(t) 
e^{[ - i \int_0^{t} dt' [\phi (t') + \kappa (t') + 
{\Delta} z^{(L)}_0 (t')] ]} , \> \>
\end{eqnarray}
and the solution of Eq. (\ref{aa5}) is given by
\begin{eqnarray}
\label{aa8}
&& \>
\>\>\>\>\>\> gz^{(L)}_2 (T) = i e^{[ i 2 \int_0^T dt' [\phi (t') + 
{\Delta} z^{(L)}_0 (t')] ]}  \\
&\times&  \int_0^T dt \mu B(t)  z^{(L)}_0 (t)  z^{(R)}_1 (t) 
e^{[ - i 2 \int_0^{t} dt' [\phi (t') +  
 {\Delta} z^{(L)}_0 (t')] ]} . \> \> \nonumber 
\end{eqnarray}

When the magnetic field is set to zero Eq. (\ref{a6})  
gives the neutrino survival probability to be 
\begin{eqnarray}
\label{aa9}
&&|\nu^{(L)}_e (T)|^2 = \nonumber \\ 
&& \exp \left\{ i \int_0^T dt \left[ 
\Delta \left( z_0^{(L)*} (t) 
-  z_0^{(L)} (t) \right) \right] \right\}. 
\end{eqnarray}
It is easy to see that this result, along with Eq. (\ref{aa3}), represents 
a resonance. Introducing
\begin{equation}
\label{aa10}
\Psi (T) = \exp \left[ - i \int_0^T dt \left( \Delta z_0^{(L)} (t) + 
\phi (t) \right) \right] 
\end{equation}
one observes that 
$|\nu^{(L)}_e (T)|^2 = |\Psi (T)|^2$. It follows from Eq. (\ref{aa3}) 
that $\Psi$ satisfies the differential equation
\begin{equation}
\label{aa11}
\frac{d^2\Psi}{d\tau^2} = - \left[ 1 + \frac{\phi^2 (t)}{\Delta^2} + 
\frac{i}{\Delta} \frac{d\phi}{d\tau} \right] \Psi.
\end{equation}
The rate of change of the probability is maximized when the right hand 
side of Eq. (\ref{aa11}) is an extremum, which is achieved when $\phi =0$. 
The width of this resonance (the MSW resonance) is $ \Gamma = 
\frac{d\phi}{dt} / \Delta$, 
which corresponds to a spatial width of $\Delta \delta r 
= 2/\Gamma$ or
\begin{equation}
\label{aa12}
\delta r_{\rm MSW} = \frac{2 \Delta}{(d\phi/dt)_{\phi=0}}.
\end{equation}
A similar argument, applied to Eq. (\ref{a4}) gives the width of the 
spin-flavor resonance to be 
\begin{equation}
\label{aa13}
\delta r_{\rm SFP} = 2 / \frac{d}{dt} \left( \frac{\phi+\kappa}{\mu B} 
\right). 
\end{equation}

As we mentioned earlier the spin-flavor precession resonance takes place 
before the MSW resonance. In most cases the quantity $z^{(L)}_0$ is
very small at the SFP resonance zone 
and can be neglected in Eqs. (\ref{aa4}) and (\ref{aa5}). In this 
approximation $z^{(L)}_2 =0$ and 
\begin{eqnarray}
\label{aa14}
&&g z^{(R)}_1 (T) = -i e^{[ i \int_0^T dt' [\phi (t') + \kappa (t') ] ]} 
\nonumber \\
&\times&  \int_0^T dt \mu B(t) 
e^{[ - i \int_0^{t} dt' [\phi (t') + \kappa (t') ] ]} , \> \>
\end{eqnarray}
Substituting these in Eq. (\ref{a6}) we obtain 
\begin{eqnarray}
\label{aa15}
&&|\nu^{(L)}_e (T)|^2 \nonumber \\ 
&=& \exp \left\{ - \left| \int_0^T dt \mu B (t) 
e^{[ i \int_0^t dt' [\phi (t') 
+ \kappa (t') ] ]} \right|^2 \right\} \nonumber \\ 
&\times&  \exp \left\{ i \int_0^T dt \left[ 
\Delta \left( z_0^{(L)*} (t) 
-  z_0^{(L)} (t) \right) \right] \right\}, 
\end{eqnarray}
or
\begin{eqnarray}
\label{aa16}
&& P (\nu_{el} \rightarrow \nu_{eL} , \mu B \neq 0) = 
P (\nu_{el} \rightarrow \nu_{eL} , \mu B = 0 ) \nonumber \\
&\times& \exp \left\{ - \left| \int_0^T dt \mu B (t) 
e^{[ i \int_0^t dt' [\phi (t') 
+ \kappa (t') ] ]} \right|^2 \right\}. 
\end{eqnarray}
If the SFP resonance width is rather small one can calculate the integral 
in Eq. (\ref{aa16}) rather accurately in the stationary phase 
approximation. The stationary point is where the derivative of the argument 
of the exponent is zero, i.e. $\phi + \kappa =0$, the SFP resonance point. 
One finally gets
\begin{eqnarray}
\label{aa17}
&& P (\nu_{el} \rightarrow \nu_{eL} , \mu B \neq 0) = 
P (\nu_{el} \rightarrow \nu_{eL} , \mu B = 0 ) \nonumber \\
&\times& \exp \left\{ - 
\frac{ 2 \pi(\mu B)^2}{|d(\phi+\kappa)/dt|_{(\phi+\kappa)=0}} \right\} .
\end{eqnarray}


\begin{thebibliography}{99}

\bibitem{Cisneros:1970nq}
A.~Cisneros,
Astrophys.\ Space Sci.\  {\bf 10}, 87 (1971).

\bibitem{Lim:1987tk}
C.~S.~Lim and W.~J.~Marciano,
Phys.\ Rev.\ D {\bf 37}, 1368 (1988).

\bibitem{Akhmedov:uk}
E.~K.~Akhmedov,
Phys.\ Lett.\ B {\bf 213}, 64 (1988).

\bibitem{Wolfenstein:1977ue}
L.~Wolfenstein,
Phys.\ Rev.\ D {\bf 17}, 2369 (1978).

\bibitem{Mikheev:wj}
S.~P.~Mikheev and A.~Y.~Smirnov,
Nuovo Cim.\ C {\bf 9}, 17 (1986).

\bibitem{Balantekin:1990jg}
A.~B.~Balantekin, P.~J.~Hatchell and F.~Loreti,
Phys.\ Rev.\ D {\bf 41}, 3583 (1990).

\bibitem{Guzzo:1998sb}
M.~M.~Guzzo and H.~Nunokawa,
Astropart.\ Phys.\  {\bf 12}, 87 (1999)
[arXiv:hep-ph/9810408].

\bibitem{Akhmedov:2002mf}
E.~K.~Akhmedov and J.~Pulido,
Phys.\ Lett.\ B {\bf 553}, 7 (2003)
[arXiv:hep-ph/0209192].

\bibitem{Friedland:2002pg}
A.~Friedland and A.~Gruzinov,
Astropart.\ Phys.\  {\bf 19}, 575 (2003)
[arXiv:hep-ph/0202095].

\bibitem{Kang:2004tx}
S.~K.~Kang and C.~S.~Kim,
Phys.\ Lett.\ B {\bf 584}, 98 (2004)
[arXiv:hep-ph/0403059].

\bibitem{Davis:cj}
R.~Davis, in {\em Proceedings of the Seventh Workshop on Grand 
Unification, Toyama, Japan}, Ed. J. Arefune (World Scientific, Singapore, 
1987), p. 237. 

\bibitem{Yoo:2003rc}
J.~Yoo {\it et al.}  [Super-Kamiokande Collaboration],
Phys.\ Rev.\ D {\bf 68}, 092002 (2003)
[arXiv:hep-ex/0307070].

\bibitem{Sturrock:2004hx}
P.~A.~Sturrock, D.~O.~Caldwell, J.~D.~Scargle, G.~Walther and M.~S.~Wheatland,
arXiv:hep-ph/0403246.

\bibitem{Raghavan:1991em}
R.~S.~Raghavan, A.~B.~Balantekin, F.~Loreti, A.~J.~Baltz, S.~Pakvasa and 
J.~Pantaleone,
Phys.\ Rev.\ D {\bf 44}, 3786 (1991).

\bibitem{Akhmedov:uk1}
E.~K.~Akhmedov,
Phys.\ Lett.\ B {\bf 255}, 84 (1991).

\bibitem{Gando:2002ub}
Y.~Gando {\it et al.}  [Super-Kamiokande Collaboration],
Phys.\ Rev.\ Lett.\  {\bf 90}, 171302 (2003)
[arXiv:hep-ex/0212067].

\bibitem{Eguchi:2003gg}
K.~Eguchi {\it et al.}  [KamLAND Collaboration],
Phys.\ Rev.\ Lett.\  {\bf 92}, 071301 (2004)
[arXiv:hep-ex/0310047].

\bibitem{Balantekin:2003jm}
A.~B.~Balantekin and H.~Yuksel,
Phys.\ Rev.\ D {\bf 68}, 113002 (2003)
[arXiv:hep-ph/0309079].

\bibitem{Bahcall:2000nu}
J.~N.~Bahcall, M.~H.~Pinsonneault and S.~Basu,
Astrophys.\ J.\  {\bf 555}, 990 (2001)
[arXiv:astro-ph/0010346].

\bibitem{Brun:1998qk}
A.~S.~Brun, S.~Turck-Chieze and P.~Morel,
Astrophys.\ J.\  {\bf 506}, 913 (1998)
[arXiv:astro-ph/9806272].

\bibitem{Adelberger:1998qm}
E.~G.~Adelberger {\it et al.},
Rev.\ Mod.\ Phys.\  {\bf 70}, 1265 (1998)
[arXiv:astro-ph/9805121].

\bibitem{Bahcall:2004fg}
J.~N.~Bahcall and M.~H.~Pinsonneault,
Phys.\ Rev.\ Lett.\  {\bf 92}, 121301 (2004)
[arXiv:astro-ph/0402114].

\bibitem{Turck}
S.~Turck-Chieze {\it et al.}, 
Astrophys.\ J.\  {\bf 555}, L69 (2001).

\bibitem{Couvidat:ba}
S.~Couvidat, S.~Turck-Chieze and A.~G.~Kosovichev,
Astrophys.\ J.\  {\bf 599}, 1434 (2003).

\bibitem{Friedland:2002is}
A.~Friedland and A.~Gruzinov,
Astrophys.\ J.\  {\bf 601}, 570 (2004)
[arXiv:astro-ph/0211377].

\bibitem{Couvidat:2003ep}
S.~Couvidat, R.~A.~Garcia, S.~Turck-Chieze, T.~Corbard, C.~J.~Henney 
and S.~Jimenez-Reyes,
Astrophys.\ J.\  {\bf 597}, L77 (2003)
[arXiv:astro-ph/0309806].

\bibitem{Antia:2000pu}
H.~M.~Antia, S.~M.~Chitre and M.~J.~Thompson,
Astron. Astrophys. {\bf 360}, 335 (2000) 
[arXiv:astro-ph/0005587].

\bibitem{bahcall}
J.N. Bahcall, {\em Neutrino Astrophysics} (Cambridge University Press, 
Cambridge, England, 1989).

\bibitem{Daraktchieva:2003dr}
Z.~Daraktchieva {\it et al.}  [MUNU Collaboration],
Phys.\ Lett.\ B {\bf 564}, 190 (2003)
[arXiv:hep-ex/0304011].

\bibitem{texono}
H.B. Li, et al, TEXONO Collaboration,
Phys. Rev. Lett. {\bf 90}, 131802 (2003).
 
\bibitem{savannah}
F.~Reines, H.S.~Gurr, and H.W. Sobel,
Phys. Rev. Lett. {\bf 37}, 315 (1976).
 

\bibitem{Vogel:iv}
P.~Vogel and J.~Engel,
Phys.\ Rev.\ D {\bf 39}, 3378 (1989).

\bibitem{kurchatov}
G.S.~Vidyakin {\it et al.},
JETP Lett.\  {\bf 55}, 206 (1992)
[Pisma Zh.\ Eksp.\ Teor.\ Fiz.\  {\bf 55}, 212 (1992)].

\bibitem{rovno}
A.~I.~Derbin, A.~V.~Chernyi, L.~A.
~Popeko, V.~N.~Muratova, G.~A.~Shishkina and S.~I.~Bakhlanov,
JETP Lett.\  {\bf 57}, 768 (1993)
[Pisma Zh.\ Eksp.\ Teor.\ Fiz.\  {\bf 57}, 755 (1993)].


\bibitem{Beacom:1999wx}
J.~F.~Beacom and P.~Vogel,
Phys.\ Rev.\ Lett.\  {\bf 83}, 5222 (1999)
[arXiv:hep-ph/9907383].


\bibitem{Liu:2004ny}
D.~W.~Liu {\it et al.}  [Super-Kamiokande Collaboration],
arXiv:hep-ex/0402015.


\bibitem{Raffelt:wa}
G.~G.~Raffelt,
``Stars as Laboratories for Fundamental Physics: The Astrophysics of
Neutrinos, Axions, and Other Weakly Interacting Particles'',
Chicago, USA: Univ. Press (1996); and references therein.

\bibitem{ioannis}
Y. Giomataris and  J.D. Vergados, hep-ex/0303045 (2003).

\bibitem{mamont} The Mamont Collaboration, Nucl. Phys. A
721 (2003) 499.

\bibitem{McLaughlin:2003yg}
G.~C.~McLaughlin and C.~Volpe,
Phys.\ Lett.\ B {\bf 591}, 229 (2004)
[arXiv:hep-ph/0312156].

\bibitem{lownu} C. Volpe, Jour. Phys. G 30 (2004) L1
 [hep-ph/0303222].


\bibitem{:2004mb}
  [KamLAND Collaboration],
arXiv:hep-ex/0406035; 
see also 
K.~Eguchi {\it et al.}  [KamLAND Collaboration],
Phys.\ Rev.\ Lett.\  {\bf 90}, 021802 (2003)
[arXiv:hep-ex/0212021].

\bibitem{Balantekin:dv}
A.~B.~Balantekin and F.~Loreti,
Phys.\ Rev.\ D {\bf 45}, 1059 (1992).

\bibitem{Balantekin:1996ag}
  A.~B.~Balantekin and J.~F.~Beacom,
  Phys.\ Rev.\ D {\bf 54}, 6323 (1996)
  [arXiv:hep-ph/9606353].

\bibitem{Miranda:2004nz}
O.~G.~Miranda, T.~I.~Rashba, A.~I.~Rez and J.~W.~F.~Valle,
arXiv:hep-ph/0406066.

\bibitem{Loreti:1994ry}
See for example F.~N.~Loreti and A.~B.~Balantekin,
Phys.\ Rev.\ D {\bf 50}, 4762 (1994)
[arXiv:nucl-th/9406003].

\bibitem{Balantekin:1996pp}
A.~B.~Balantekin, J.~M.~Fetter and F.~N.~Loreti,
Phys.\ Rev.\ D {\bf 54}, 3941 (1996)
[arXiv:astro-ph/9604061].

\bibitem{Friedland:2000rn}
A.~Friedland,
Phys.\ Rev.\ D {\bf 64}, 013008 (2001)
[arXiv:hep-ph/0010231].


\end{thebibliography}
\end{document}